\documentclass[a4paper,12pt]{article}
\usepackage{amsmath}
\usepackage{accents}
\usepackage{bm}
\usepackage[english]{babel}
\usepackage{graphicx}
\usepackage[usenames]{color}
\usepackage{colortbl}
\usepackage[font=footnotesize,figurewithin=none]{caption}
\usepackage{subcaption}
\usepackage{floatrow}
\usepackage[toc,page,title]{appendix}

\begin{document}

\centerline {\LARGE{Derivation of the robustness from the concurrence}}
\medskip
\centerline {Yu. S. Krynytskyi$^1$, A. R. Kuzmak$^2$}
\centerline {\small \it E-Mail: yurikryn@gmail.com$^1$, andrijkuzmak@gmail.com$^2$}
\medskip
\centerline {\small \it Department for Theoretical Physics, Ivan Franko National University of Lviv,}
\medskip
\centerline {\small \it 12 Drahomanov St., Lviv, UA-79005, Ukraine}
\medskip

{\small

Adding the maximally mixed state with some weight to the entanglement system leads to disentanglement of the latter. For each predefined entangled state
there exists a minimal value of this weight for which the system loses its entanglement properties. This values was proposed to use as a quantitative measure
of entanglement called robustness [G. Vidal, R. Tarrach, Phys. Rev. A {\bf 59}, 141 (1999)]. Using the concurrence, we propose the derivation of this measure for the system of two qubit. Namely, for a two-qubit pure state an exact expression of robustness is obtained. Finally, in the same way, the robustness
of special cases of mixed two-qubit states is calculated.

\medskip

{\bf Key words}: robustness; concurrence; mixed state; entangled state; two-qubit system.

\medskip

}

\section{Introduction \label{sec1}}

Entanglement is a property which appears only for quantum-mechanical system \cite{horodecki2009}. It plays a crucial role in quantum-mechanical processes and
is important for implementation of quantum-information algorithms. Testing Bell's inequality \cite{bell1964} for entangled states of photons,
Aspect et al. experimentally solved the EPR paradox \cite{aspect1982}. The simplest scheme of quantum teleportation of the qubit state \cite{bennet1993},
which was experimentally realized by Zeilinger's group \cite{zeilinger1997}, requires the preparation of a two-qubit entangled state as a quantum channel.
The efficiency of quantum computers \cite{feynman1982,kane1998} is due to algorithms based on the preparation of quantum entangled states
(see, for instance, \cite{barenco1995,divincenzo2000,kitaev2002,krokhmalskii2004}). In \cite{giovannetti20031,giovannetti20032} it was shown that presence
of entanglement in a quantum system enhances the speed of evolution of this system. Thus, the evolution of the system through entangled states
occurs more rapidly than through disentangled ones. This property is important for optimization of quantum calculations. The connection
between the degree of entanglement and time of evolution was also studied in papers \cite{zander2007,borras20081,borras20082,zhao2009}.

So, the efficiency of the implementation of the above-mentioned processes depends on the degree of entanglement in a system.
Therefore, it is necessary to determine whether a state is entangled and to quantify the value of its entanglement.
For the first time the quantative study of entanglement states was made in papers \cite{popescu1995,horodecki1996}. The authors examined the behavior of Bell
inequalities \cite{bell1964} on the mixed entangled states. It was shown that there is a family
of the entangled mixed states, which does not violate the Bell inequalities. The first step in finding the criteria that define the distinction between
entangled and separable state was done by Peres \cite{peres1996} and Horodecki family \cite{horodecki1997,horodecki1996b}.
Lewenstein and Sanpera was proposed an algorithm that allows to find the best separable approximation to an arbitrary density matrix \cite{lewenstein1998}.
This algorithm is based on fact that an arbitrary density matrix $\rho$ can be expressed as follows $\rho=\omega\rho_S+(1-\omega)P_e$, where $\rho_s$ is a
separable density matrix for which the weight $\omega$ is maximal, $P_e=\vert\psi_e\rangle\langle\psi_e\vert$ is a projector on an pure entangled state
$\vert\psi_e\rangle$, and $\omega\in[0,1]$. The state $\rho_s$ is also called the
best separable approximation of $\rho$. In \cite{karnas2001} it was shown  that always there exist a unique such decomposition that $\omega$
is maximall. Further this algorithm was considered for different kinds of states \cite{karnas2001,kraus2000,karnas&lewenstein2001,wellens2001,quesada2014}.

For finding the quantitative distinction between the entangled and disentangled states the different measures of entanglement were proposed.
The most basic measure for the bipartite system is called the entanglement of formation, which is defined as follows \cite{bennet19961}
\begin{eqnarray}
E(\rho)=\min\sum_ip_iE(\vert\psi_i\rangle).
\label{entropyofent}
\end{eqnarray}
This is the average entanglement of the pure states $\vert\psi_i\rangle$ of the decomposition minimized over all decompositions of mixed states $\rho$ with
probabilities $p_i$, which satisfies condition $\sum_ip_i=1$. Here the entanglement of the pure state is defined as an entropy of either of the two subsystem
$A$ and $B$ \cite{bennet19962,popescu1997}
\begin{eqnarray}
E(\vert\psi\rangle)=-\mathrm{Tr} \left(\rho_A\log_2\rho_A\right)=-\mathrm{Tr}\left(\rho_B\log_2\rho_B\right),
\label{entropyofentpure}
\end{eqnarray}
where $\rho_{A(B)}$ is the partial trace of $\vert\psi\rangle\langle\psi\vert$ over the subsystem $B(A)$. In \cite{wootters1998}, Wootters presented an explicit
formula for finding the value of entanglement (concurrence $C$) of a two-qubit state. He obtained the formula for entanglement of formation (\ref{entropyofent}) of a mixed state $\rho$ of two qubits
as a monotonic function of concurrence
\begin{eqnarray}
&&E(\rho)=-\frac{1+\sqrt{1-C(\rho)^2}}{2}\log_2\left(\frac{1+\sqrt{1-C(\rho)^2}}{2}\right)\nonumber\\
&&-\frac{1-\sqrt{1-C(\rho)^2}}{2}\log_2\left(\frac{1-\sqrt{1-C(\rho)^2}}{2}\right).
\label{entof formation}
\end{eqnarray}
Thus, the value of entanglement of a particular two-qubit state defined by density matrix $\rho$ can be calculated as follows
\begin{eqnarray}
C(\rho)=\max\{0,\lambda_1-\lambda_2-\lambda_3-\lambda_4\}.
\label{wootters}
\end{eqnarray}
Here, $\lambda_i$ are the eigenvalues, in decreasing order, of the Hermitian matrix $R=\sqrt{\sqrt{\rho}\tilde{\rho}\sqrt{\rho}}$, where
$\tilde{\rho}=\left(\sigma_y\otimes\sigma_y\right)\rho^*\left(\sigma_y\otimes\sigma_y\right)$. Note that $\lambda_i$ are real and positive numbers.
For calculations it is convenient to use the eigenvalues of the non-Hermitian matrix $\rho\tilde{\rho}$ which have the form $\lambda_i^2$. The value of concurrence ranges from 0 for separable states to 1
for maximally entangled states. Another interesting definition of the measure of entanglement is the geometric measure of entanglement proposed by Shimony \cite{shimony1998}.
Its properties and different definitions were considered in papers \cite{brody2001,wei2003} and \cite{chen2014}, respectively.
In paper \cite{frydryszak2017}, the geometrical measure was quantified in terms of mean values of observables of the entangled system. An algorithm to
explore entanglement of a bipartite system based on maximization of the Schmidt norms was discussed in paper \cite{reuvers2018}.

In papers \cite{vidal1999,steiner2003} were proposed to quantify the measure of entanglement in terms of mixing. Adding the maximally mixed state with a certain weight to the entangled state
leads to its factorization. The minimal value of weight that transforms the system into a disentangled state can be used as a degree of entanglement of
this system (robustness). The robustness of state $\rho_0$ relative to separable state $\rho_s$ is the minimal value of $\omega\geq 0$ for which $\rho=(1-\omega)\rho_0+\omega\rho_s$ is separable \cite{vidal1999}.
In the present paper, we apply this consideration to a two-qubit system (section \ref{secdeff}). As a result, in section \ref{secpqs},
using the concurrence as a quantitative measure of entanglement, we obtain an explicit expression for the robustness relative to the maximally mixed state in the case of pure states.
Also, in section \ref{secmqs} we derive the robustness for some special cases of mixed states. Finally, the discussion is presented in section \ref{concl}.

\section{Space of a two-qubit quantum state and definition of robustness \label{secdeff}}

An arbitrary two-qubit quantum state $\rho_0$ can be expressed as the decomposition of the pure states
\begin{eqnarray}
\rho_0=\sum_i p_i\rho_i,
\label{mixedstate}
\end{eqnarray}
where
\begin{eqnarray}
\rho_i=\vert\psi_i\rangle\langle\psi_i\vert
\label{mixedstatest}
\end{eqnarray}
is the density matrix of the pure state $\vert\psi_i\rangle$ with probability $p_i$.  State (\ref{mixedstate}) is defined by fifteen real parameters.
Therefore, the space which contains all two-qubit quantum states is 15-dimensional. This space is bounded to the 14-dimensional convex region
which contains all pure states and mixed states of rank-2 and 3. Inside the region there are the mixed states with rank-4 containing the maximally mixed state
\begin{eqnarray}
\rho_{\max}=\frac{1}{4}I,
\label{maxmixst}
\end{eqnarray}
where $I$ is the unit matrix. This region is divided into domain that contains entangled states and convex domain that contains disentangled states.
The transition from any entangled state to disentangled state can be accomplished by adding the maximally mixed state (\ref{maxmixst})
with the weight coefficient $\omega\in[0,1]$ as follows
\begin{eqnarray}
\rho=(1-\omega)\rho_{0}+\omega\rho_{\max}.
\label{rhodisent}
\end{eqnarray}
For each predefined entangled state $\rho_{0}$ there exists the minimal weight $\omega_c$ for which state (\ref{rhodisent}) becomes disentangled. The value of $\omega_c$ depends on
the degree of the entanglement of state $\rho_{0}$. The lower degree of the entanglement of $\rho_{0}$, the smaller value takes the $\omega_c$, and
vice versa. So, this parameter can be used as a quantitative measure of entanglement of quantum states and it is called robustness \cite{vidal1999}. Note, that this is the definition of robustness of state $\rho_0$
relative to the maximally mixed state $\rho_{max}$. It is worth noting that such considerations are valid for
any many-body quantum system. In the following section, using the concurrence as quantitative measure of entanglement, we obtain the exact expression for $\omega_c$ in the case of a pure quantum state of two qubits.
For a mixed state it is difficult to obtain a general expression for $\omega_c$. Therefore, we consider some special cases of mixed states.

\section{Entanglement of a pure quantum state \label{secpqs}}

In general, an arbitrary quantum state of two qubits can be represented by the Schmidt decomposition
\begin{eqnarray}
\vert\psi\rangle=c_1\vert\alpha_1\rangle\vert\beta_1\rangle+c_2 e^{i\chi}\vert\alpha_2\rangle\vert\beta_2\rangle,
\label{purestatesd}
\end{eqnarray}
where $\vert\alpha_1\rangle$, $\vert\alpha_2\rangle$ are the orthogonal states which belong to the first qubit
\begin{eqnarray}
\vert\alpha_1\rangle=\frac{\vert\uparrow\rangle+\alpha\vert\downarrow\rangle}{\sqrt{1+\vert\alpha\vert^2}},\quad \vert\alpha_2\rangle=\frac{\alpha^*\vert\uparrow\rangle-\vert\downarrow\rangle}{\sqrt{1+\vert\alpha\vert^2}},\nonumber
\end{eqnarray}
and $\vert\beta_1\rangle$, $\vert\beta_2\rangle$ are the orthogonal states which belong to the second qubit
\begin{eqnarray}
\vert\beta_1\rangle=\frac{\vert\uparrow\rangle+\beta\vert\downarrow\rangle}{\sqrt{1+\vert\beta\vert^2}},\quad \vert\beta_2\rangle=\frac{\beta^*\vert\uparrow\rangle-\vert\downarrow\rangle}{\sqrt{1+\vert\beta\vert^2}}.\nonumber
\end{eqnarray}
Here $\alpha$, $\beta$ are some complex parameters and $\chi$ is some real parameter. The Schmidt coefficients $c_1$ and $c_2$ are real and positive satisfying
the normalization condition $c_1^2+c_2^2=1$. Representation (\ref{purestatesd}) is useful for calculating measure of entanglement between qubits. Indeed,
using concurrence (\ref{wootters}) with state (\ref{purestatesd}) we obtain that its value of entanglement is defined only by the Schmidt coefficients
as follows
\begin{eqnarray}
C(\vert\psi\rangle)=2c_1c_2.
\label{concurrenceps}
\end{eqnarray}
It achieves the maximal value ($C=1$) when $c_1=c_2=1/\sqrt{2}$. Note that making the local unitary transformations
with the first and second qubits in state (\ref{purestatesd}) we can reach state \cite{KTTSM}
\begin{eqnarray}
\vert\psi\rangle=c_1\vert\uparrow\uparrow\rangle+c_2\vert\downarrow\downarrow\rangle.
\label{bell}
\end{eqnarray}
These transformations do not change the entanglement of the system. Therefore, for further calculations of entanglement we use state (\ref{bell}).

Let us study the influence of the unit matrix on the value of entanglement of state (\ref{bell}). For this purpose we construct the
density matrix of this state and add maximally mixed state (\ref{maxmixst}) to it. As a result, the state of the system becomes mixed. In the basis
spanned by the states $\vert\uparrow\uparrow\rangle$ and $\vert\downarrow\downarrow\rangle$ the density matrix of this state takes the form
\begin{eqnarray}
\rho=\left( \begin{array}{ccccc}
(1-\omega)c_1^2+\frac{\omega}{4} & 0 & 0  & (1-\omega)c_1c_2\\
0 & \frac{\omega}{4} & 0 & 0 \\
0 & 0 & \frac{\omega}{4} & 0 \\
(1-\omega)c_1c_2 & 0 & 0 & (1-\omega)c_2^2+\frac{\omega}{4}
\end{array}\right).
\label{densitymatrix}
\end{eqnarray}
Here $\omega$ defines the degree of mixing of quantum state. The value of $\omega=1$ corresponds to the
maximally mixed state. Using definition of concurrence (\ref{wootters}) let us calculate the degree of entanglement for state (\ref{densitymatrix})
(see appendix \ref{appa})
\begin{eqnarray}
C(\rho)=\max\{0,\left(1-\omega\right)2c_1c_2-\omega/2\}.
\label{woottersps}
\end{eqnarray}
As we can see that concurrence of mixed state (\ref{densitymatrix}) contains concurrence (\ref{concurrenceps}) of pure state (\ref{purestatesd}).
We obtain the critical (minimal) value of $\omega$, for which pure state (\ref{bell}) becomes disentangled, when we equate to zero expression
$\left(1-\omega\right)2c_1c_2-\omega/2$ and then solve it with respect to $\omega$.
As a result we obtain the robustness of two-qubit pure state
\begin{eqnarray}
\omega_c=\frac{C(\vert\psi\rangle)}{C(\vert\psi\rangle)+1/2},
\label{criticalomega}
\end{eqnarray}
where $C(\vert\psi\rangle)$ is defined by equation (\ref{concurrenceps}). This expression takes the maximal value $2/3$ for $c_1=c_2=1/\sqrt{2}$, which corresponds
to the maximally entangled pure state (\ref{purestatesd}), and minimal value $0$ for $c_1=0$ or $c_2=0$, which corresponds to the disentangled
pure state. The value of $\omega_c$ indicates the "amount" of unit matrix that must be added to pure state (\ref{purestatesd})
with predefined $c_1$ and $c_2$ in order to disentangle it. So, the value of $\omega_c$ can be used as a measure of entanglement of pure quantum state. Then value $\omega_c=0$ corresponds
to the disentangled state and $\omega_c=2/3$ corresponds to maximally entangled state. Note that if the initial state is spanned
by the basis vectors $\vert\uparrow\uparrow\rangle$, $\vert\uparrow\downarrow\rangle$, $\vert\downarrow\uparrow\rangle$ and $\vert\downarrow\downarrow\rangle$
as follows $a\vert\uparrow\uparrow\rangle+b\vert\uparrow\downarrow\rangle+c\vert\downarrow\uparrow\rangle+d\vert\downarrow\downarrow\rangle$
then $c_1c_2$ in formula (\ref{criticalomega}) should be changed into $\vert ad-bc\vert$. Here $a$, $b$, $c$ and $d$ are complex parameters
which satisfy the normalization condition $\vert a\vert^2+\vert b\vert^2+\vert c\vert^2+\vert d\vert^2=1$.

\section{Entanglement of a mixed quantum state \label{secmqs}}

For an arbitrary two-qubit mixed state, using the definition of concurrence, it is difficult to obtain a general expression of robustness $\omega_c$. Therefore, in this section we
consider few special cases of mixed states.

\subsection{Rank-2 mixed state}

First of all, let us study the entanglement of {\it rank-2 mixed state} $\rho_0$ with density matrix (\ref{mixedstate}), where state $\vert\psi_i\rangle$
is given on subspace spanned by vectors $\vert\uparrow\uparrow\rangle$, $\vert\downarrow\downarrow\rangle$ as follows
\begin{eqnarray}
\vert\psi_i\rangle=c_{1i}\vert\uparrow\uparrow\rangle+c_{2i}e^{i\chi_i}\vert\downarrow\downarrow\rangle.
\label{pqsranktwo}
\end{eqnarray}
Here $c_{ij}$, $\chi_i$ are real coefficients which satisfy the normalization condition $c_{1i}^2+c_{2i}^2=1$, and the number of states in ensemble
can be arbitrary. The mixture of this state with maximally mixed one (\ref{maxmixst}) takes the form
\begin{eqnarray}
\rho=\left( \begin{array}{ccccc}
(1-\omega)\sum_i p_ic_{1i}^2+\frac{\omega}{4} & 0 & 0  & (1-\omega)\sum_i p_ic_{1i}c_{2i}e^{-i\chi_i}\\
0 & \frac{\omega}{4} & 0 & 0 \\
0 & 0 & \frac{\omega}{4} & 0 \\
(1-\omega)\sum_i p_ic_{1i}c_{2i}e^{i\chi_i} & 0 & 0 & (1-\omega)\sum_i p_ic_{2i}^2+\frac{\omega}{4}
\end{array}\right).
\label{densitymatrixmixssubs}
\end{eqnarray}
By making the same calculations as in the previous case (see appendix \ref{appa}) we obtain that the minimal value of weigh $\omega_c$ which transforms
state (\ref{densitymatrixmixssubs}) into disentangled one has the same form as in case of pure state (\ref{criticalomega}), hovewer,
one should replace $C(\vert\psi\rangle)$ with the following expression
\begin{eqnarray}
C(\rho_0)=2\left[\left(\sum_i p_ic_{1i}c_{2i}\cos\chi_i\right)^2+\left(\sum_i p_ic_{1i}c_{2i}\sin\chi_i\right)^2\right]^{1/2}.
\label{wootterstworank}
\end{eqnarray}
This expression defines the concurrence of state (\ref{densitymatrixmixssubs}) with $\omega=0$. As we can see, this state takes the maximal
value of entanglement if $c_{1i}=c_{2i}=1/\sqrt{2}$ and if all the $\chi_i$ are equal to each other. Note that this result is valid in the case of mixed states which are defined
on the subspace spanned by vectors $\vert\uparrow\downarrow\rangle$, $\vert\downarrow\uparrow\rangle$.

\subsection{$X$-state}

Now let us consider a two-qubit mixed state of rank-4 called {\it an X-state} \cite{ali2010,mendonca2014}. Note that such a state can be generated
by two interacting spins-$1/2$ which evolve under the influence of the thermal bath \cite{ying-hua2012,kuzmak2019}. To study the connection of the concurrence and robustness of the $X$-state
we prepare the mixture of the above-discussed rank-2 mixed states
spanned by $\vert\uparrow\uparrow\rangle$, $\vert\downarrow\downarrow\rangle$ and $\vert\uparrow\downarrow\rangle$, $\vert\downarrow\uparrow\rangle$ vectors,
respectively. The density matrix of this mixture with maximally mixed state (\ref{maxmixst}) takes the form
\begin{eqnarray}
\rho=\left( \begin{array}{ccccc}
(1-\omega)\sum_i p_ic_{1i}^2+\frac{\omega}{4} & 0 & 0  & (1-\omega)\sum_i p_ic_{1i}c_{2i}e^{-i\chi_i}\\
0 & 0 & 0 & 0 \\
0 & 0 & 0 & 0 \\
(1-\omega)\sum_i p_ic_{1i}c_{2i}e^{i\chi_i} & 0 & 0 & (1-\omega)\sum_i p_ic_{2i}^2+\frac{\omega}{4}
\end{array}\right)\nonumber\\
+\left( \begin{array}{ccccc}
0 & 0 & 0  & 0\\
0 & (1-\omega)\sum_i q_id_{1i}^2+\frac{\omega}{4} & (1-\omega)\sum_i q_id_{1i}d_{2i}e^{-i\phi_i} & 0 \\
0 & (1-\omega)\sum_i q_id_{1i}d_{2i}e^{i\phi_i} & (1-\omega)\sum_i q_id_{2i}^2+\frac{\omega}{4} & 0 \\
0 & 0 & 0 & 0
\end{array}\right),
\label{densitymatrixmix}
\end{eqnarray}
where $q_i$ and $d_{ij}$, $\phi_i$ are the probabilities and parameters which define the part of state spanned by vectors
$\vert\uparrow\downarrow\rangle$, $\vert\downarrow\uparrow\rangle$. Here, parameters $d_{ij}$ satisfy the normalization condition $d_{1i}^2+d_{2i}^2=1$,
and for probabilities we have the following condition $\sum_ip_i+\sum_iq_i=1$.
It is worth noting that this case includes a rank-3 state if we leave only one component of the density matrix that corresponds to the first or second
subspace, respectively.

Constructing for state (\ref{densitymatrixmix}) the $R$ matrix defined after expression (\ref{wootters}) we obtain its eigenvalues
\begin{eqnarray}
&&\lambda_{1,2}=\left[(1-\omega)^2A^2+\frac{\omega}{4}(1-\omega)P+\left(\frac{\omega}{4}\right)^2\right]^{1/2}\pm (1-\omega)F,\nonumber\\
&&\lambda_{3,4}=\left[(1-\omega)^2B^2+\frac{\omega}{4}(1-\omega)Q+\left(\frac{\omega}{4}\right)^2\right]^{1/2}\pm (1-\omega)G,
\label{lambdmixs}
\end{eqnarray}
where
\begin{eqnarray}
&&A=\left[\sum_{i,j}p_ip_jc_{1i}^2c_{2j}^2\right]^{1/2},\quad B=\left[\sum_{i,j}q_iq_jd_{1i}^2d_{2j}^2\right]^{1/2},\nonumber\\
&&F=\left[\left(\sum_ip_ic_{1i}c_{2i}\cos\chi_i\right)^2+\left(\sum_ip_ic_{1i}c_{2i}\sin\chi_i\right)^2\right]^{1/2},\nonumber\\
&&G=\left[\left(\sum_iq_id_{1i}d_{2i}\cos\phi_i\right)^2+\left(\sum_iq_id_{1i}d_{2i}\sin\phi_i\right)^2\right]^{1/2},\nonumber
\end{eqnarray}
\begin{eqnarray}
P=\sum_ip_i,\quad Q=\sum_iq_i.\nonumber
\label{compoflambda}
\end{eqnarray}
Since $\lambda_1\geq\lambda_2$ and $\lambda_3\geq\lambda_4$, and eigenvalues $\lambda_{1,2}$ and $\lambda_{3,4}$ are symmetric between themselves,
it is enough to obtain the expression for $\omega_c$ in the case of $\lambda_1\geq\lambda_3$. Then the concurrence has the form
\begin{eqnarray}
C(\rho)=\max\left\{0,2(1-\omega)F-2\left[(1-\omega)^2B^2+\frac{\omega}{4}(1-\omega)Q+\left(\frac{\omega}{4}\right)^2\right]^{1/2}\right\}.
\label{concurrrank4}
\end{eqnarray}
For an entangled state the value in curly brackets is positive, and robustness $\omega_c$ is the following
\begin{eqnarray}
\omega_c=\frac{8(F^2-B^2)+Q-\sqrt{Q^2+4(F^2-B^2)}}{8(F^2-B^2)+2Q-1/2}.
\label{criticalomegamix}
\end{eqnarray}
Note that in the opposite case of $\lambda_3\geq\lambda_1$ one should make the following replacemets in this expression:
$F\rightarrow G$, $B\rightarrow A$ and $Q\rightarrow P$. Value (\ref{criticalomegamix}) vanishes when $F=B$, and increases when $B$ tends to zero.
For the case of $Q=0$ expression (\ref{criticalomegamix}) turns into expression for rank-2 state (\ref{criticalomega}) with $C(\rho_0)$ defined
by formula (\ref{wootterstworank}). It is worth noting that the entanglement of mixture of mixed states with the same entanglement and which
belong to different subspaces always equals $0$. This fact is easy to check if we put $\lambda_1=\lambda_3$
and $\lambda_2=\lambda_4$ in Wooters equation (\ref{wootters}).

So, the entanglement of the mixture of states from subspaces defined by vectors $\vert\uparrow\uparrow\rangle$, $\vert\downarrow\downarrow\rangle$ and
$\vert\uparrow\downarrow\rangle$, $\vert\downarrow\uparrow\rangle$ is always less than the more entangled component of this mixture. Moreover, if we mix
the states with the same entanglement from these subspaces then we obtain the disentangled state.

\subsection{Rank-4 mixed state with real parameters}

In this subsection we consider a more general case, namely, a two-qubit rank-4 mixed state with real parameters. The density matrix of this state with maximally mixed state (\ref{maxmixst}) reads
\begin{eqnarray}
\rho=\left( \begin{array}{ccccc}
(1-\omega)a+\frac{\omega}{4} & (1-\omega)b & -(1-\omega)b  & (1-\omega)d\\
(1-\omega)b & (1-\omega)c+\frac{\omega}{4} & (1-\omega)f & (1-\omega)b \\
-(1-\omega)b & (1-\omega)f & (1-\omega)c+\frac{\omega}{4} & -(1-\omega)b \\
(1-\omega)d & (1-\omega)b & -(1-\omega)b & (1-\omega)a+\frac{\omega}{4}
\end{array}\right),
\label{densitymatrixreal}
\end{eqnarray}
where $a$, $b$, $c$, $d$ and $f$ are real parameters. Here, parameters $a$ and $c$ satisfy condition $2a+2c=1$. It  is easy to check that in this case matrix $\tilde{\rho}$ coincides with matrix $\rho$.
This means that the matrix $R=\rho$ and eigenvalues $\lambda_i$ are the eigenvalues of the density matrix $\rho$. The eigenvalues of this matrix are the following
\begin{eqnarray}
&&\lambda_{1}=(1-\omega)\left(a-d\right)+\frac{\omega}{4},\quad \lambda_{2}=(1-\omega)\left(c+f\right)+\frac{\omega}{4},\nonumber\\
&&\lambda_{3,4}=\frac{1-\omega}{2}\left(a+d+c-f\pm\sqrt{(a+d-c+f)^2+16b^2}\right)+\frac{\omega}{4}.
\label{lambdmixsreal}
\end{eqnarray}
Since $\lambda_3\geq\lambda_{1}$ and $\lambda_3\geq\lambda_{4}$, we have two cases of concurrence (\ref{wootters}) for state (\ref{densitymatrixreal}): $\lambda_2\geq\lambda_3$ and $\lambda_3\geq\lambda_2$.
Let us condider them in detail.\\
\begin{itemize}
\item $\lambda_2\geq\lambda_3$\\
In this case the concurrence takes the form
\begin{eqnarray}
C(\rho)={\rm max}\left\{0,2(1-\omega)(f-a)-\frac{\omega}{2}\right\}.
\label{conccasea}
\end{eqnarray}
As we can see, the state remains an entangled if $f>a$. Then the robustness $\omega_c$ reads
\begin{eqnarray}
\omega_c=\frac{2(f-a)}{2(f-a)+1/2}.
\label{robcasea}
\end{eqnarray}
\item $\lambda_3\geq\lambda_2$\\
Here, the concurrence takes the form
\begin{eqnarray}
C(\rho)={\rm max}\left\{0,(1-\omega)\left(\sqrt{\left((2a+d+f-\frac{1}{2}\right)^2+16b^2}+d-f-\frac{1}{2}\right)-\frac{\omega}{2}\right\}.
\label{conccaseb}
\end{eqnarray}
From this expression we obtain the robustness in the form
\begin{eqnarray}
\omega_c=1-\frac{1}{2\left(\sqrt{(2a+d+f-1/2)^2+16b^2}+d-f\right)}.
\label{robcaseb}
\end{eqnarray}
\end{itemize}

As an example, let us consider the mixture of two pure states\linebreak
$\vert\psi_1\rangle=1/2\left(\vert\uparrow\uparrow\rangle+\vert\uparrow\downarrow\rangle-\vert\downarrow\uparrow\rangle+\vert\downarrow\downarrow\rangle\right)$
and $\vert\psi_2\rangle=1/2\left(\vert\uparrow\uparrow\rangle-\vert\uparrow\downarrow\rangle+\vert\downarrow\uparrow\rangle+\vert\downarrow\downarrow\rangle\right)$.
The density matrix $\rho_0=p\vert\psi_1\rangle\langle\psi_1\vert+(1-p)\vert\psi_2\rangle\langle\psi_2\vert$ of this mixture has the real parameters $a=c=d=1/4$, $f=-1/4$ and $b=(2p-1)/4$.
Then the eigenvalues of density matrix (\ref{densitymatrixreal}) read $\lambda_1=\lambda_2=\omega/4$ and $\lambda_{3,4}=(1-\omega)(1\pm \vert 2p-1\vert)/2+\omega/4$.
Since $\lambda_3$ is the largest eigenvalue we use equations (\ref{conccaseb}), (\ref{robcaseb}) and obtain $C\left(\rho\right)={\rm max}\left(0,(1-\omega)\vert 2p-1 \vert -\omega/2\right)$,
$\omega_c=\vert 2p-1\vert/(\vert 2p-1\vert+1/2)$. As we can see, the maximal value of the entanglement corresponds to pure states ($p=0,1$), and the minimal value of the entanglement corresponds to the mixed state with $p=1/2$.

\section{Discussion \label{concl}}

Adding the maximally mixed state with some weight to the density matrix of entangled system leads to its disentanglement. For each
entangled state there exists a critical (minimal) value of this weight for which the system ceases to be entangled. Moreover, for a more entangled
state this value is greater and vice versa. This weight was proposed to use as a quantitative measure of entanglement and it is called robustness \cite{vidal1999}. Using the definition of concurrence, we have derived these measure
for a two-qubit system. Namely, explicit expression (\ref{criticalomega}) of this measure for any predefined two-qubit pure state was obtained.
It was shown that if the state is maximally entangled then the two-thirds of the maximally mixed state should be added to the pure state to transform it into
a disentangled state. Therefore, this measure takes the values between $2/3$ for maximally entangled states and $0$ for factorized ones. In the case of mixed states,
we were not able to obtain a general expression for this weight. However, we considered the special cases of mixed states. For a
rank-$2$ mixed state spanned by vectors $\vert\uparrow\uparrow\rangle$, $\vert\downarrow\downarrow\rangle$ or
$\vert\uparrow\downarrow\rangle$, $\vert\downarrow\uparrow\rangle$ it was shown that the robustness is determined by expression
(\ref{criticalomega}) with (\ref{wootterstworank}). In addition, we mixed the states from these two subspaces and obtained robustness in the form
(\ref{criticalomegamix}). As a result we showed that the entanglement of such a mixture is always less than the more entangled component of it.
Finally, we consider a rank-4 mixed state which consists of real parameters. The value of the entanglement of this state is defined by its eigenvalues.
Depending on the ratio between these eigenvalues we obtain two expressions to determine robustness (\ref{robcasea}) and (\ref{robcaseb}).

\section{Acknowledgements}

We are grateful to Profs. Volodymyr Tkachuk and Andrij Rovenchak for helpful discussions. This work was partly supported by Project 77/02.2020 (No.~0120U104801) from National Research Foundation of Ukraine.

\begin{appendices}
\section{Derivation of $\omega_c$ for pure two-qubit quantum state \label{appa}}
\setcounter{equation}{0}
\renewcommand{\theequation}{A\arabic{equation}}

In this appendix using Wootters definition of concurrence (\ref{wootters}) we obtain the degree of entanglement of state (\ref{densitymatrix}).
In this case matrix $\tilde{\rho}$ has the form
\begin{eqnarray}
\tilde{\rho}=\left(\sigma_y\otimes\sigma_y\right)\rho^*\left(\sigma_y\otimes\sigma_y\right)=\left( \begin{array}{ccccc}
(1-\omega)c_2^2+\frac{\omega}{4} & 0 & 0  & (1-\omega)c_1c_2\\
0 & \frac{\omega}{4} & 0 & 0 \\
0 & 0 & \frac{\omega}{4} & 0 \\
(1-\omega)c_1c_2 & 0 & 0 & (1-\omega)c_1^2+\frac{\omega}{4}
\end{array}\right).\nonumber\\
\label{tilderho}
\end{eqnarray}
The matrix $\rho\tilde{\rho}$ can be expressed as follows
\begin{eqnarray}
\scriptsize{
\rho\tilde{\rho}=\left( \begin{array}{ccccc}
2(1-\omega)^2c_1^2c_2^2+\frac{\omega}{4}\left(1-\omega\right)+\left(\frac{\omega}{4}\right)^2 & 0 & 0  & 2\left((1-\omega)c_1^2+\frac{\omega}{4}\right)(1-\omega)c_1c_2\\
0 & \left(\frac{\omega}{4}\right)^2 & 0 & 0 \\
0 & 0 & \left(\frac{\omega}{4}\right)^2 & 0 \\
2\left((1-\omega)c_2^2+\frac{\omega}{4}\right)(1-\omega)c_1c_2 & 0 & 0 & 2(1-\omega)^2c_1^2c_2^2+\frac{\omega}{4}\left(1-\omega\right)+\left(\frac{\omega}{4}\right)^2
\end{array}\right).}\nonumber\\
\label{rhotilderho}
\end{eqnarray}
Then the eigenvalues $\lambda_i^2$ of this matrix satisfy the following equations
\begin{eqnarray}
&&\lambda^4-2\lambda^2\left[2(1-\omega)^2c_1^2c_2^2+\frac{\omega}{4}-3\left(\frac{\omega}{4}\right)^2\right]+\left(\frac{\omega}{4}-3\left(\frac{\omega}{4}\right)^2\right)^2=0,\nonumber\\
&&\left(\lambda^2-\left(\frac{\omega}{4}\right)^2\right)^2=0.
\label{equtionforlambda}
\end{eqnarray}
Solving these equations we take into account only the positive solutions and write them in decreasing order
\begin{eqnarray}
&&\lambda_1=\sqrt{(1-\omega)^2c_1^2c_2^2+\frac{\omega}{4}-3\left(\frac{\omega}{4}\right)^2}+(1-\omega)c_1c_2\nonumber\\
&&\lambda_2=\sqrt{(1-\omega)^2c_1^2c_2^2+\frac{\omega}{4}-3\left(\frac{\omega}{4}\right)^2}-(1-\omega)c_1c_2\nonumber\\
&&\lambda_{3,4}=\frac{\omega}{4}.
\label{lambdas}
\end{eqnarray}
Now, substituting these eigenvalues in formula (\ref{wootters}) we obtain expression (\ref{woottersps}). Then equating to zero expression
$\left(1-\omega\right)2c_1c_2-\omega/2$ and solving it with respect to $\omega$ we obtain formula (\ref{criticalomega}).

\end{appendices}

\end{document}